\documentclass{PoS}

\title{Modeling Hybrid Stars}

\ShortTitle{Modeling Hybrid Stars}

\author{Veronica Dexheimer\\
       UFSC, Florianopolis, BR and \\
       Gettysburg College, Gettysburg, USA\\
       E-mail: \email{vantoche@gettysburg.edu}}

\author{Stefan Schramm\\
        FIAS, Johann Wolfgang Goethe University, Frankfurt, DE\\
        E-mail: \email{schramm@th.physik.uni-frankfurt.de}}

\author{\speaker{Jirina Stone}\\
        University of Oxford, Oxford, UK and\\
        University of Tennessee, Knoxville, USA\\
        E-mail: \email{jstone12@utk.edu}}

\abstract{We study the so called hybrid stars, which are hadronic stars that contain a core of deconfined quarks. For this purpose, we make use of an extended version of the SU(3) chiral model. Within this approach, the degrees of freedom change naturally from hadrons (baryon octet) to quarks (u, d, s) as the temperature and/or density increases. At zero temperature we are still able to reproduce massive stars, even with the inclusion of hyperons.}

\FullConference{XII International Symposium on Nuclei in the Cosmos\\
August 5-12, 2012\\
Cairns, Australia}

\begin{document}

\section{Introduction and Model Description}

In the center of compact stars the density can reach several times the nuclear saturation density. At such extreme conditions, the baryon chemical potential is high enough for nucleons to be converted into hyperons. In addition, QCD calculations have shown that those densities are high enough for the hadrons to be deconfined into quarks \cite{2010PhRvD..81j5021K}. In this work we are going to consider that neutron stars are actually hybrid stars composed of hadrons surrounding a core of quark matter.

To describe these stars, we are going to use an extended version of the hadronic SU(3) chiral model that also contains quarks \cite{Dexheimer:2009hi}. We are further going to use the mean-field approximation in which all particles contribute to the global mean-field interactions and are in turn affected by them. In this model the order parameters $\sigma$ and $\Phi$ signal chiral symmetry restoration and quark deconfinement, respectively. Both transitions are expected to happen at high densities and/or temperatures.

\begin{figure}
  \center
  \includegraphics[trim=0cm 0cm 0cm .07cm, clip=true,height=.3\textheight]{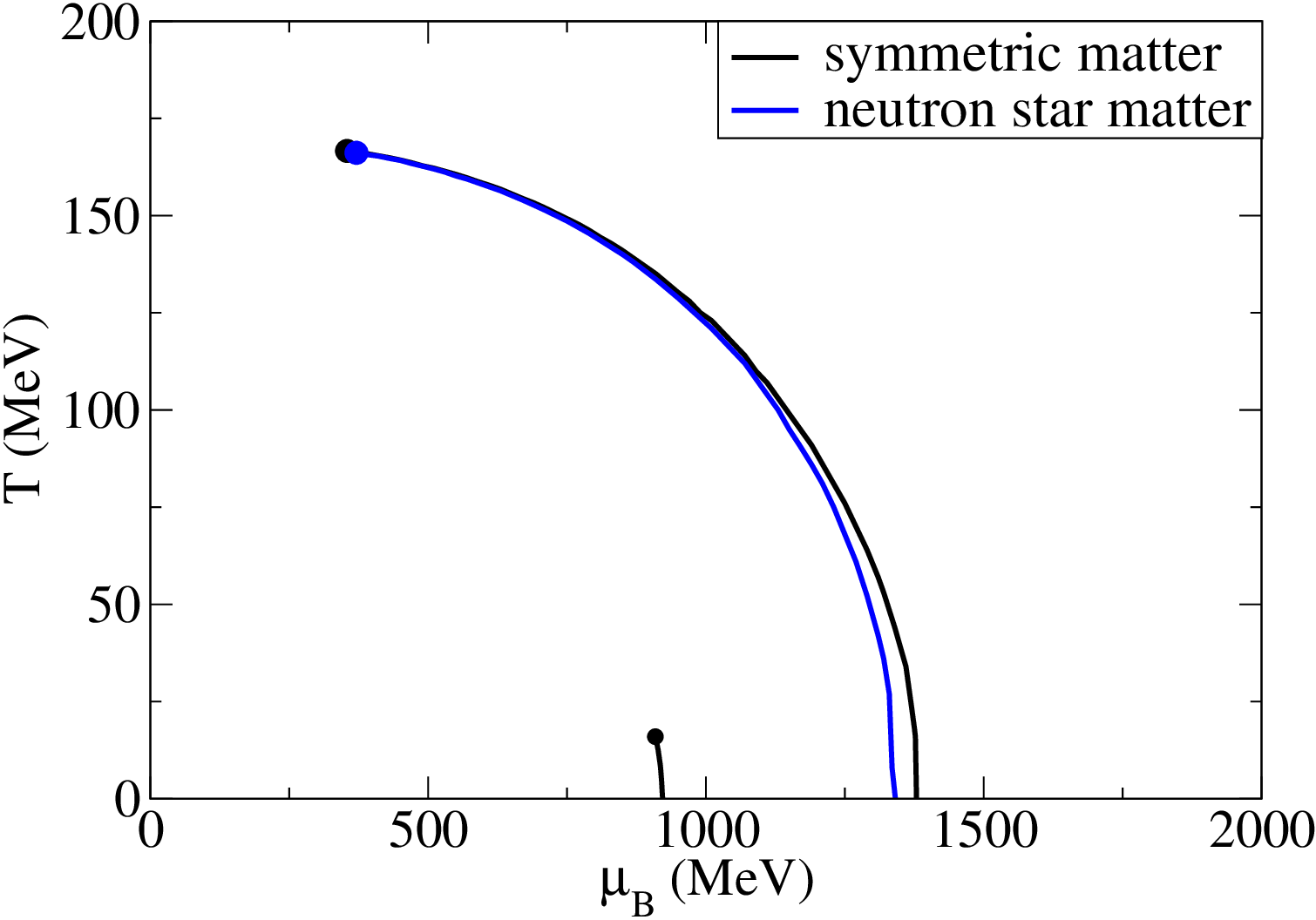}
  \caption{QCD Phase Diagram: Temperature versus baryon chemical potential}
  \label{phase}
\end{figure}

Fig.~\ref{phase} shows a phase diagram obtained from analyzing the behavior of the order parameters. The first order phase transition line ends in a critical point beyond which the transition becomes a crossover in accordance with lattice QCD constraints \cite{Fodor:2004nz}. The different lines correspond to symmetric matter (same number of protons and neutrons) and neutron star matter (in beta equilibrium and charge neutral). For the first case, the nuclear matter liquid-gas phase transition is also shown. The potential for the deconfinement order parameter $\Phi$ is an extension of the Polyakov loop potential \cite{Roessner:2006xn} modified to also depend on baryon chemical potential. In this way the model is able to describe the entire QCD phase diagram, including the low and zero temperature part which is relevant for compact star physics.

Within the model described above, the baryon and quark masses are generated by the scalar mesons, whose mean-field values correspond to the isoscalar ($\sigma$) and isovector ($\delta$) light quark-antiquark condensates as well as the strange quark-antiquark condensate ($\zeta$). In addition, there is a small explicit mass term $M_0$ and the term containing $\Phi$
\begin{equation} 
M_{B}^*=g_{B\sigma}\sigma+g_{B\delta}\tau_3\delta+g_{B\zeta}\zeta+M_{0_B}+g_{B\Phi} \Phi^2,
\label{1}
\end{equation}
\begin{equation}
M_{q}^*=g_{q\sigma}\sigma+g_{q\delta}\tau_3\delta+g_{q\zeta}\zeta+M_{0_q}+g_{q\Phi}(1-\Phi),
\label{2}
\end{equation}
where the coupling constant values can be found in Ref.~\cite{Dexheimer:2009hi}.

\begin{figure}
  \center
  \includegraphics[trim=0cm 0cm 0cm .07cm, clip=true,height=.3\textheight]{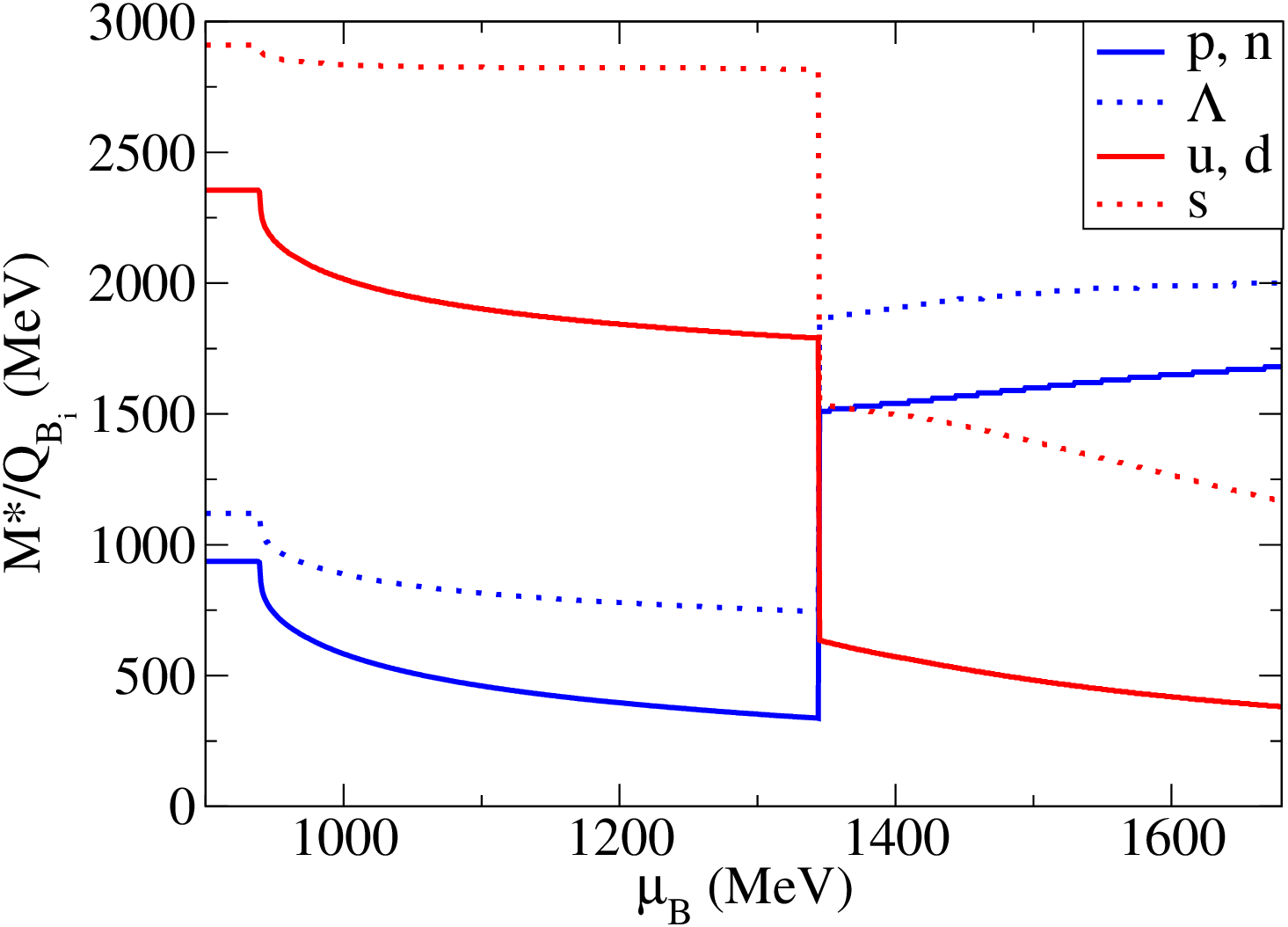}
  \caption{Normalized effective mass versus baryon chemical potential for star matter at zero temperature.}
  \label{meff}
\end{figure}

With the increase of density/temperature, the $\sigma$ field (non-strange chiral condensate) decreases from its high value at zero density, causing the effective masses of the particles to decrease towards chiral symmetry restoration. The field $\Phi$ assumes non-zero values with the increase of temperature/density and, due to its presence in the baryons effective mass, suppresses their presence. On the other hand, the presence of the $\Phi$ field in the effective mass of the quarks, included with a negative sign, ensures that they will not be present at low temperatures/densities.
This can be clearly seen at Fig.~\ref{meff} where the normalized effective mass (quark effective masses are multiplied by $3$) is shown for star matter at zero temperature.

\section{Results and Conclusions}

\begin{figure}
  \center
  \includegraphics[trim=0cm 0cm 0cm .07cm, clip=true,height=.3\textheight]{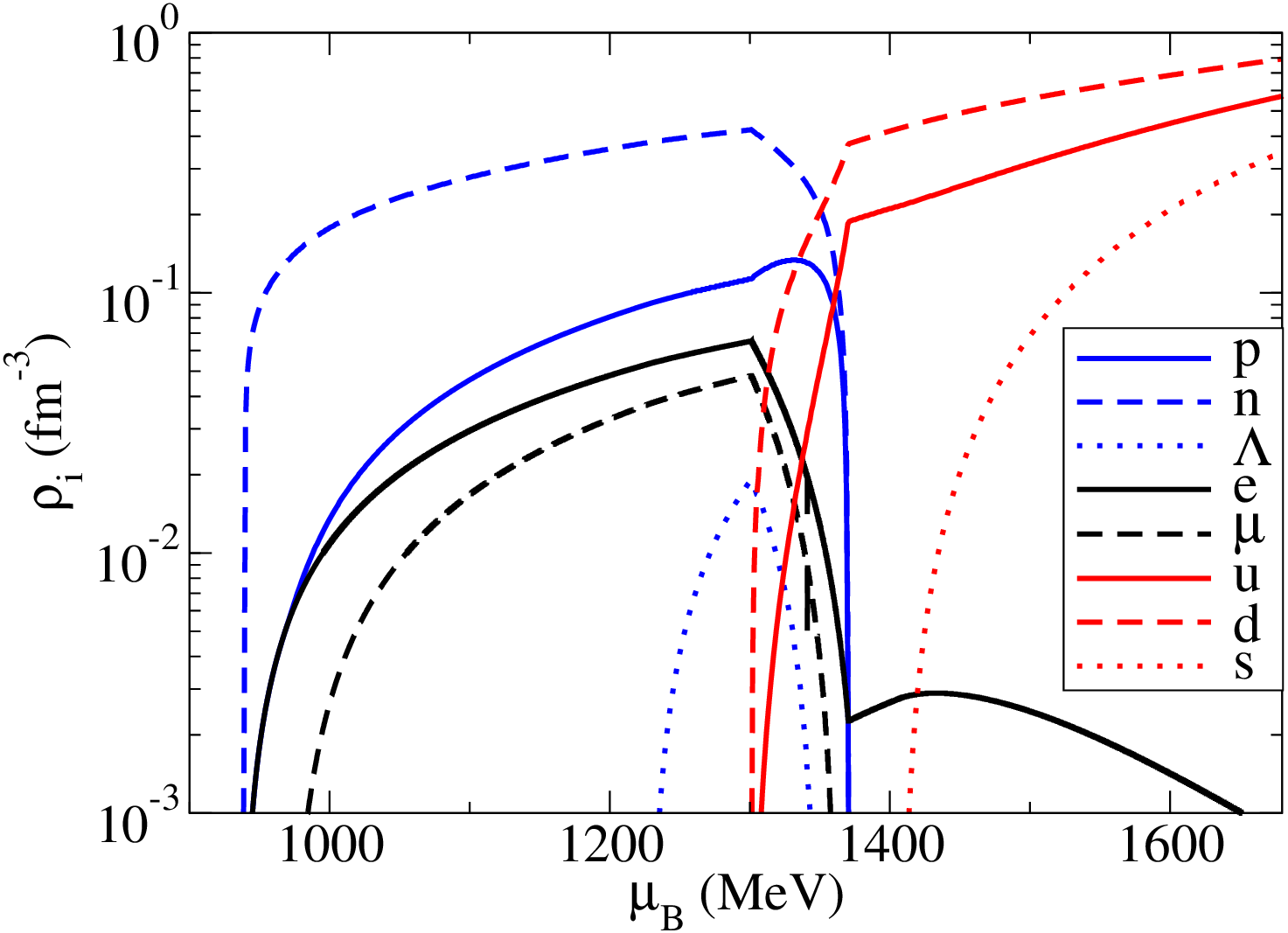}
  \caption{Particle population versus baryon chemical potential for star matter at zero temperature.}
  \label{pop}
\end{figure}

For more realistic star calculations we assume that charge neutrality is globally conserved. This means that hadronic and quark phases don't necessarily have to be charge neutral when separated, but can also be charge neutral when combined. As a result, a mixed phase appears. Fig.~\ref{pop} shows the particle population for star matter at $T=0$. As the baryon chemical potential increases, the neutrons convert to protons and electrons (for charge neutrality), followed by muons and Lambdas. All the other hyperons are suppressed by the appearance of the quarks. First up and down quarks slowly appear and later the strange quark. The amount of electrons present in the system is substantial in the hadronic phase, but not in the quark phase, as the down quarks take care of balancing the positive charged particles.

\begin{figure}
  \center
  \includegraphics[trim=0cm 0cm 0cm .07cm, clip=true,height=.3\textheight]{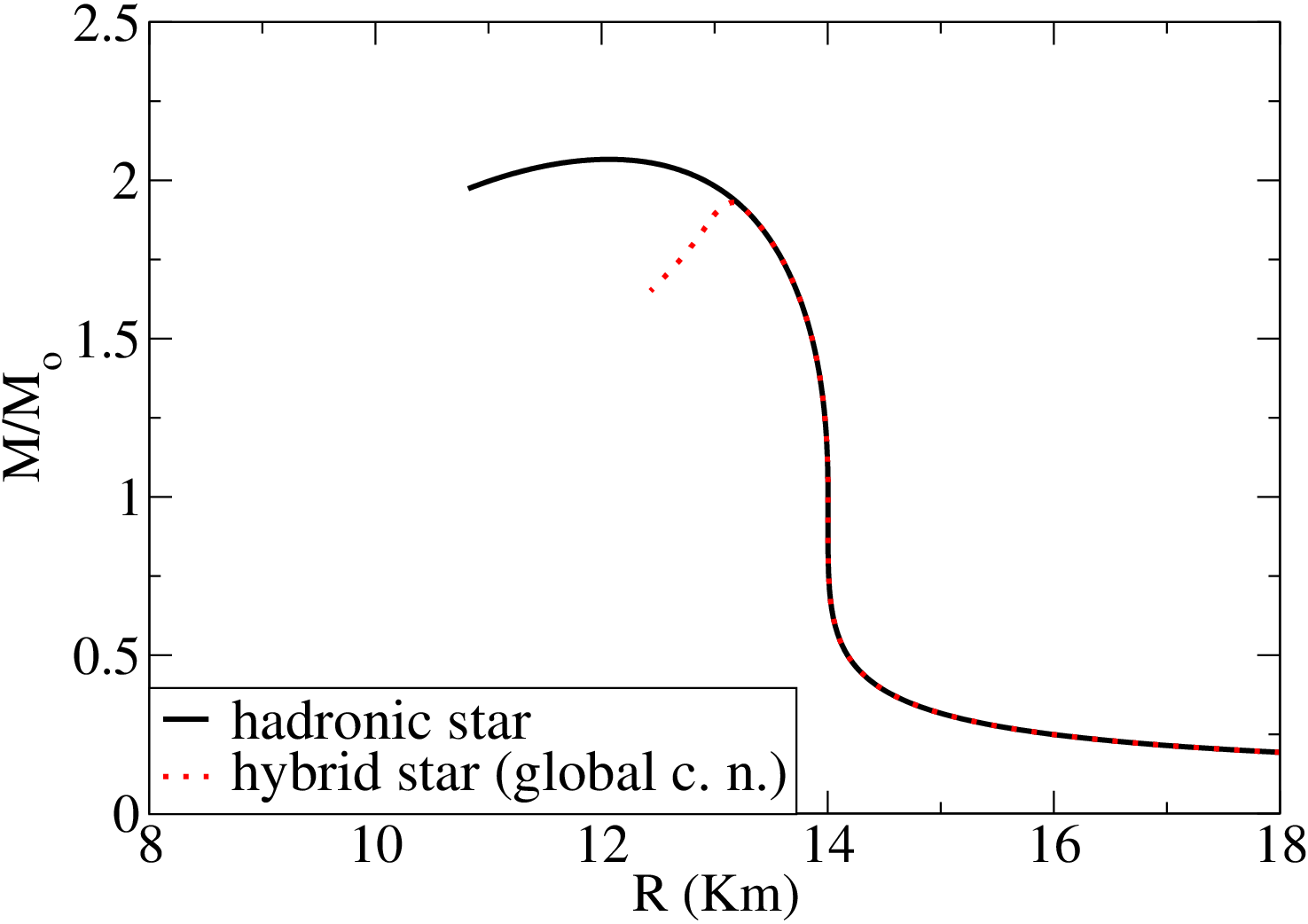}
  \caption{Mass-Radius Diagram: Star mass versus radius for different central densities (for star matter at zero temperature).}
  \label{mass}
\end{figure}

Finally, we plug in our equation of state into the Tolman Oppenheimer Volkoff equations \cite{Tolman:1939jz,Oppenheimer:1939ne}, the solution of Einstein's equations for the specific case of spherical, isotropic, static stars. For each possible central density we obtain a respective star mass and radius as shown in Fig.~\ref{mass}. The two different lines represent pure hadronic and hybrid stars. Within this model pure quark stars are not stable, only pure hadronic or hybrid stars. I the case of hybrid stars, they can contain up to $2$ km of "mixed" matter.

In conclusion, our model is suitable for the description of neutron stars. We predict stars as massive as the most massive pulsar observed (PSR J16142230 with a mass of $1.97\pm0.04$ M$\odot$). The predicted radii also lie in the allowed range being practically the same for hadronic or hybrid stars. We were also able to be in good agreement with data concerning the cooling behavior of hybrid stars \cite{Negreiros:2010hk} and heavy ion collision data \cite{Steinheimer:2009nn}.

A major advantage of our work compared to other studies of hybrid stars is that because we have only one equation of state for different degrees of freedom we can study in detail the way in which chiral symmetry is restored and the way deconfinement occurs at high temperature/density. The model additionally shows a realistic structure of the phase transition over the whole range of chemical potentials and temperatures as well as phenomenologically acceptable results for saturated nuclear matter.


\section{Acknowledgments}

V. D. acknowledges support from CNPq-Brazil.


\bibliographystyle{utphys}
\bibliography{skeleton}

\end{document}